# Highly sensitive spin-flop transition in antiferromagnetic van-der Waals material $M$PS$_3$ ($M$ = Ni and Mn)

Rabindra Basnet[1], Aaron Wegner[1], Krishna Pandey[2], Stephen Storment[1], Jin Hu[1,2*]

[1]Department of Physics, University of Arkansas, Fayetteville, Arkansas 72701, USA

[2]Materials Science and Engineering Program, Institute for Nanoscience and Engineering, University of Arkansas, Fayetteville, Arkansas 72701, USA

## Abstract

Recent developments in two-dimensional (2D) magnetism have motivated the search for novel van-der Waals (vdWs) magnetic materials to explore new magnetic phenomenon in the 2D limit. Metal thiophosphates, $M$P$X_3$, is a class of magnetic vdWs materials with antiferromagnetic (AFM) ordering persisting down to the atomically thin limit. The magnetism in this material family has been found to be highly dependent on the choice of transition metal $M$. In this work, we have synthesized the intermediate compounds Ni$_{1-x}$Mn$_x$PS$_3$ ($0 \leq x \leq 1$) and investigated their magnetic properties. Our study reveals that the variation of Ni and Mn content in Ni$_{1-x}$Mn$_x$PS$_3$ can efficiently tune the spin-flop transition, likely due to the modulation of the magnetic anisotropy. Such effective tunning offers a promising candidate to engineer 2D magnetism for future device applications.

[*] jinhu@uark.edu

Recent breakthroughs in two-dimensional (2D) magnetic materials open a new route in exploring intrinsic magnetism in the 2D limit. The discoveries of novel 2D magnets [1–6] provide opportunities not only to understand the mechanism of low dimensional magnetism but also to design next-generation devices. For example, advances in devices and heterostructures involving 2D magnets enable the effective tuning of their electronic and magnetic properties, providing a fascinating platform to study fundamental physics [7–18]. These breakthroughs have greatly enriched our understanding of magnetism in the 2D limit.

Among 2D magnetic materials, one model system is metal thiophosphates $M$P$X_3$ ($M$ = transition metal ions, $X$ = chalcogen ions). $M$P$X_3$ materials crystallize in a monoclinic layered structure with $C_{2/m}$ space group, in which the transition metal ions form a honeycomb layer and carry localized magnetic moments [19–22]. The van der Waals (vdW)-type crystal structure allows for the inter-layer intercalation [23–27] and mechanical exfoliations down to atomically thin layers [5,28–36]. The magnetic and electronic properties of the $M$P$X_3$ family are strongly dependent on the choice of the transition metal $M$. MnPS$_3$, NiPS$_3$, and FePS$_3$ have been found to exhibit Heisenberg-type, $XY$ or $XXZ$-type, and Ising-type antiferromagnetism, respectively [37–42]. The Neel temperature ($T_N$) also varies with transition metal $M$, increasing from 78 K for MnPS$_3$, to 118, 122, and 155 K for FePS$_3$, CoPS$_3$, and NiPS$_3$, respectively [38,40,42–45]. Besides magnetism, the electronic properties of $M$P$X_3$ are also tunable with $M$, exhibiting a wide range of band gaps from 1.3 eV for FePSe$_3$ to 3.5 eV for ZnPS$_3$ [22]. Recent studies on NiPS$_3$ and MnPS$_3$ have established these materials as platforms to study correlated electrons in 2D magnetic materials [32,46,47]. Furthermore, theoretical and experimental high-pressure studies revealed insulator-to-metal transitions in MnPS$_3$ [47,48], MnPSe$_3$ [48], FePS$_3$ [49–52], NiPS$_3$ [47,53] and V$_{0.9}$PS$_3$ [51,54], and even the emergence of superconductivity in FePSe$_3$ [55].

The structural similarity of the $M$P$X_3$ materials allows for the synthesis of polymetallic "mixed" compounds with substitution of transition metal $M$ [56–66]. Varying lattice constant and magnetic moments through elemental substitution is an effective way to tune magnetism and probe the underlying physics of magnetic materials. Coupling of magnetism with lattice constants and symmetries in $M$P$X_3$ has been theoretically predicted [67]. Tuning of magnetism has also been observed in mixed compounds such as Mn$_x$Fe$_{1-x}$PS$_3$, Mn$_{1-x}$Fe$_x$PSe$_3$, Fe$_{0.5}$Ni$_{0.5}$PS$_3$ and Mn$_{1-x}$Zn$_x$PS$_3$ [56–66]. For example, magnetic order and spin orientations in MnPS$_3$ can be systematically tuned by Zn substitution for Mn [56–58]. Spin glass is also found to arise from the competing 3$d$ magnetism in Mn substituted FePS$_3$ [59–61].

In this work, we report the systematic study on previously unexplored Ni$_{1-x}$Mn$_x$PS$_3$ ($0 \leq x \leq 1$). We synthesized single crystals and characterized the evolution of magnetic properties with substitutions. In addition to the known field-induced spin-flop (SF) transition in MnPS$_3$, we have also discovered the previously unreported SF transitions in NiPS$_3$. Furthermore, we found these SF transitions are extremely sensitive to magnetic substitutions, likely due to the re-orientation of the magnetic moments of 3$d$ elements controlled by trigonal distortion of $M$S$_6$ octahedral. Such tunable magnetism offers a promising platform for studying new phenomenon arising from 2D magnetism and future device applications.

Single crystals of Ni$_{1-x}$Mn$_x$PS$_3$ ($0 \leq x \leq 1$) used in this work were synthesized by the chemical vapor transport method using I$_2$ as the transport agent. Elemental powders with stoichiometric ratio were sealed in a quartz tube and placed in a two-zone furnace with a temperature gradient from 750 to 550 °C for 1 week. Millimeter to centimeter size single crystals with various colors have been obtained, as shown in the insets of Fig. 1a. These crystals are thin plate-like and easily exfoliable with hexagonal facets which is consistent with the vdW structure of *M*P*X*$_3$ shown in Fig. 1b. The elemental compositions examined by energy-dispersive x-ray spectroscopy (EDS) reveal successful Ni-Mn substitution. We have carefully characterized the compositions of all the Ni$_{1-x}$Mn$_x$PS$_3$ single crystals used in this paper. The Mn contents *x* throughout this paper are measured values. In addition to composition analysis, the x-ray diffraction (XRD) on single crystals also indicates successful substitution. As shown in Fig. 1 (a), the (*00L*) diffraction peaks exhibit a systematic low-angle shift with increasing *x* in Ni$_{1-x}$Mn$_x$PS$_3$, which is consistent with the greater ionic radius of Mn$^{2+}$ as compared with Ni$^{2+}$, respectively.

In Fig. 2 we present the magnetic property characterizations for pristine MnPS$_3$ and NiPS$_3$. For MnPS$_3$, below the Néel temperature $T_N \sim 78$ K, a spin-flop (SF) transition can be observed in the isothermal magnetization measured under out-of-plane magnetic field (*H*⊥ab) [Fig. 2(a)] but is absent under in-plane field (*H*//*ab*) [Fig. 2(b)], which is consistent with the previous reports [5,34,65,68,69]. Although the SF transition is largely explored in both bulk and atomically thin MnPS$_3$, it has not been *directly* discovered in other members of this family except FePS$_3$, in which a metamagnetic transition occurs at very high field ($\mu_0 H > 35$ T) [70]. Here we report the first discovery of SF transition in NiPS$_3$. Unlike in MnPS$_3$ in which the SF transition occurs when *H*⊥ab, the metamagnetic transition in NiPS$_3$ takes place with in-plane field, which is characterized by a clear upturn in magnetization above a critical spin-flop field $\mu_0 H_{SF} \approx 6$ T below the Néel temperature $T_N \approx 155$ K, as shown in Figs. 2(d) and 2(e). The need for relatively high magnetic field could be the reason that prevented the discovery of SF transition in earlier studies [38,71]. On the other hand, the recent magneto-photoluminescence experiment implies a spin re-orientation in bulk NiPS$_3$ under a much higher in-plane magnetic field of 15 T [32]. Such a high critical field, which is 2.5 times higher than that in our magnetization measurements, might be attributed to the nature of photoluminescence as an indirect probe.

In MnPS$_3$ and NiPS$_3$, the SF transition occurring under different field directions is in line with their magnetic structures: The Mn moments in MnPS$_3$ are aligned along out-of-plane direction [58,69], while the Ni moments in NiPS$_3$ mostly lie within the plane [31,38,71], as shown in the insets of Figs. 2b and 2c, respectively. In collinear AFM systems, a magnetic field along the easy axis exceeding a critical spin-flop field $H_{SF}$ forces the magnetic moments to rotate [72,73]. In such a SF state, the moments re-orient themselves to a canted configuration perpendicular to the field direction, resulting in a net moment along the easy axis [72,73]. Therefore, the SF transition behaves differently in MnPS$_3$ and NiPS$_3$, as illustrated in Figs. 2g and 2h. Furthermore, because the in-plane projection of Ni moments in NiPS$_3$ forms a collinear AFM structure along the *a*-axis (Fig. 2c, lower inset), the SF transition is expected to show in-plane anisotropy. To examine this, we measured magnetization with the magnetic field applied along or perpendicular to the hexagonal edges of a NiPS$_3$ single crystal, as shown in Fig. 2(e). Indeed, magnetization and SF transition are found to be dependent on in-plane field-orientations.

The scenario of SF transition in NiPS$_3$ is also supported by the temperature dependence of magnetic susceptibility. As shown in Fig. 2(f), susceptibility measured with in-plane field ($\chi_{\parallel}$) displays a low temperature upturn, which becomes more significant at higher fields ($\mu_0 H > 5$ T). A similar low temperature upturn has been observed in MnPS$_3$ [65]. In addition, the zero-field cooling (ZFC, solid lines) and field cooling (FC, dashed lines) data display weak but clear irreversibility above 5 T, which also becomes more visible at higher fields. The development of low temperature upturn and irreversibility can be understood in terms of the ferromagnetic component from the uncompensated canted moments along the easy axis in the SF state, as illustrated in Fig. 2h. This irreversibility disappears at 140 K when applying 9 T field, coinciding well with the temperature above which the SF transition vanishes as seen in the isothermal magnetization [Fig. 2(d), inset]. Such 140 K "disappearing temperature" is lower than the magnetic ordering temperature $T_N \approx 155$ K, which can be attributed to the fact that a greater field is needed for SF transition at higher temperatures, particularly when approaching $T_N$. $H_{SF}$ enhancement upon increasing temperature is widely seen in other SF systems [5,74–78], which can be interpreted using the molecular field theory [73,78]: In a weakly anisotropic antiferromagnet, spin-flop field $H_{SF}$ can be estimated by $(H_{SF})^2 = 2K/(\chi_\perp - \chi_\parallel)$, where $K$ is the anisotropy constant, $\chi_\perp$ and $\chi_\parallel$ are the perpendicular and parallel susceptibilities, respectively [58,73,78,79]. Generally, in AFM systems the difference between $\chi_\perp$ and $\chi_\parallel$ reduces more quickly than the magnetic anisotropy constant $K$ upon increasing temperature, leading to enhanced $H_{SF}$ [78].

The distinct SF transitions in NiPS$_3$ and MnPS$_3$ due to their different magnetic structures (Figs. 2g and 2h) motivate us to further study the "mixed" compounds Ni$_{1-x}$Mn$_x$PS$_3$. As shown in Fig. 3, the magnetism is highly tunable with Ni-Mn substitution. Under both in-plane ($H\|ab$, Fig. 3a) and out-of-plane ($H\perp ab$, Fig. 3b) magnetic fields, the magnetization exhibits a systematic enhancement with increasing Mn content $x$, which is consistent with the much larger magnetization of the pristine MnPS$_3$ than that of NiPS$_3$ [Figs. 2(a-d)] and can be ascribed to greater magnetic moment of Mn$^{2+}$ than Ni$^{2+}$. Interestingly, we found that the SF transitions are extremely sensitive to Ni-Mn substitution. The SF transition in NiPS$_3$ under in-plane field disappears with 5% Mn substitution (*i.e.*, $x = 0.05$ in Ni$_{1-x}$Mn$_x$PS$_3$). Similarly, under out-of-plane field, $H_{SF}$ in MnPS$_3$ is reduced by half with 5% Ni substitution (*i.e.*, $x = 0.95$) and disappears upon 10% substitution ($x = 0.9$).

Further insights can be gained from the careful comparison between the isothermal magnetizations. In Fig. 3d we show isothermal magnetization $M(H)$ of NiPS$_3$ (*i.e.*, $x = 0$) and 5% Mn-substituted (*i.e.*, $x = 0.05$) samples, reproduced from Figs. 3a and 3b. With Mn substitution, in addition to the enhancement of magnetization and the absence of SF transition as mentioned above, another interesting behavior is the sublinear field dependence at low fields. Such nonlinear $M(H)$ is more pronounced under in-plane field [Inset, Fig. 3(d)], *i.e.*, the magnetic field direction that SF transition occurs for NiPS$_3$. Similarly, at the MnPS$_3$ side, 5% Ni-substitution ($x = 0.95$) also introduces remarkable low field nonlinearity in $M(H)$ under out-of-plane field, as shown in the inset of Fig. 3e. Such nonlinearity has been verified with multiple samples and careful removal of background signal from the sample holder. The observed low field sublinear $M(H)$ in lightly substituted compounds is in sharp contrast with the linear $M(H)$ in pristine NiPS$_3$ and MnPS$_3$, implying the development of ferromagnetic component with substitution. Such behavior has also been observed in Zn-substituted MnPS$_3$. The previous study [58] on Zn substitution for Mn in MnPS$_3$ suggests the breakdown of long-range magnetic order due to non-magnetic impurities in local substituted regions, which leads to "weakly bound" Mn moments. The polarization of these Mn moments causes the low-field nonlinear $M(H)$. Another study [57] also propose that for MnPS$_3$ in which the dipolar anisotropy dominates, the local dipole field is along the out-of-plane direction. As the consequence, when Mn-magnetism is diluted by Zn substitution, the absence of the magnetic moment of one Mn site would affect the closest Mn in the neighboring layers and cause their magnetic moments to be canted, leading to an average staggered magnetic moment in a larger scale [57]. Similarly, the nonlinear $M(H)$ at low fields in our Ni$_{1-x}$Mn$_x$PS$_3$ samples can likely be attributed to the polarization of canted moment originating from substitution. This scenario is further supported by the temperature dependence of susceptibility measurements. As shown in Fig.

4a, consistent with the nonlinear $M(H)$ seen under $H\|ab$ in the $x = 0.05$ sample (Fig. 3d, inset), the in-plane susceptibility ($\chi_\|$) displays a clear upturn at low temperatures that suggests the development of a ferromagnetic component due to moment canting. Likewise, close to the MnPS$_3$ side, for $x = 0.95$ sample in which the nonlinear $M(H)$ is significant (Fig. 3e, inset), a susceptibility upturn is also seen, as shown in Fig. 4c. The upturn become more obvious with increasing the Ni content to 10% ($x = 0.9$), as shown in Fig. 4b.

The tunable SF transition and the presence of a ferromagnetic component by Ni-Mn substitution is expected because of the distinct magnetic structures in NiPS$_3$ and MnPS$_3$. However, the observed high sensitivity to light substitution is surprising. For example, only 5% Ni substitution can substantially reduce $H_{SF}$ by half in MnPS$_3$, which is much less than the amount of Zn needed (~20%) to reduce $H_{SF}$ by the same amount [58,65]. The $H_{SF}$ suppression by non-magnetic Zn substitution has been ascribed to the reduction of magnetic anisotropy with magnetic dilution [58]. The Ni-Mn substitution in our Ni$_{1-x}$Mn$_x$PS$_3$, however, induces magnetic impurities. Hence, it is necessary to consider the magnetic interaction to understand the observed sensitive doping dependence in Ni$_{1-x}$Mn$_x$PS$_3$. In $M$P$X_3$ compounds, magnetism has been found to be related to a structural distortion [23,80–83]. As illustrated in Fig. 3c, every metal atom $M$ in $M$P$X_3$ is located at the center of an octahedron formed by six $X$ atoms. Such $MX_6$ octahedra, however, possess a trigonal distortion that is characterized by the angle $\theta$ between the trigonal axis (perpendicular to the $ab$-plane) and the $M$-S bond. Therefore, magnetism in $M$P$X_3$ can be described by an isotropic Heisenberg Hamiltonian with additional single-ion anisotropy due to the combined effect of crystal field and spin-orbit splitting: $H = -2\sum JS_iS_j + DS_{iz}^2$, where $J$ and $D$ are the exchange and crystal field parameters, respectively [37]. The relative strength of $J$ and $D$ leads to various magnetic structures in $M$P$X_3$, so the trigonal distortion plays critical role in determining the nature of magnetic interactions in $M$P$X_3$ [37,80]. In NiPS$_3$ and MnPS$_3$, $\theta$ has been found to be 51.05° and 51.67°, respectively [80]. Compared with $\theta \approx 54.75°$ for the undistorted octahedra, the greater trigonal distortion in NiPS$_3$ leads to much stronger single-ion anisotropy (0.3 meV) than that of MnPS$_3$ (0.0086 meV), as revealed by inelastic scattering measurements [42,80,81]. This causes the Ni moments to be aligned within the basal plane perpendicular to the trigonal axis, with a small out-of-plane component likely due to weak dipolar anisotropy, as shown in the insets of Figs. 2c [37,43,82]. In contrast, the effect of crystal field and spin-orbit splitting is found to be negligible for the high spin ground state of $Mn^{2+}$ in MnPS$_3$ [37], so the magnetism of the less distorted MnPS$_3$ is governed by the dipolar anisotropy that leads to out-of-plane moment direction with small tilt towards $a$-axis [42,69,82], as shown in the inset of Fig. 2b. Therefore, the tuning of trigonal distortion by Ni-Mn substitution would significantly affect the magnetic isotropy and further efficiently modify the SF transitions. A similar mechanism has also been proposed for the suppression of $H_{SF}$ in MnPS$_3$ under pressure [82].

In addition to magnetic anisotropy, magnetic exchange interaction may also play an important role. The magnetism in $M$P$X_3$ is mainly mediated through superexchange interaction, thereby affected by the $M$-$X$-$M$ bonding angle [37,81]. Additionally, the $d$-electron occupancy of $M$ is important in determining the sign and the nature of the superexchange [37,80,81]. The 3$d$ orbitals of $M^{2+}$ ion consist of high energy $e_g$ and low energy $t_{2g}$ groups, and their occupancies determine the strength of the exchange interactions [80,83]. This has recently been demonstrated by the inelastic neutron scattering measurements which reveals reduced exchange interactions in CoPS$_3$ as compared to NiPS$_3$ due to distinct occupancy of $t_{2g}$ orbital [80]. A similar scenario should also occur in Ni$_{1-x}$Mn$_x$PS$_3$, in which the Ni-Mn substitution modifies the magnetic exchange. Indeed, the magnitudes of all the exchanges, except the weakest second nearest-neighbor interaction ($J_2$), are found to systematically increase with decreasing spin on $M^{2+}$ ion and increasing $M$-S-$M$ bond angles from MnPS$_3$ to NiPS$_3$ [80,81,83].

Among the magnetic anisotropy and magnetic exchange interactions, the former may govern the SF transitions in Ni$_{1-x}$Mn$_x$PS$_3$. The magnetic ordering temperature be related to the strength of the magnetic interaction. We have extracted $T_N$ for Ni$_{1-x}$Mn$_x$PS$_3$ from magnetic susceptibility and heat capacity measurements, as summarized in Fig. 4 (e). For both NiPS$_3$ and MnPS$_3$, $T_N$ decreases with substitution until reaching to a minimum value of ~12 K in $x = 0.5$ sample. A similar trend has also been observed in other mixed $MPX_3$ systems such as Mn$_{1-x}$Zn$_x$PS$_3$ [65], Mn$_{1-x}$Fe$_x$PS$_3$ [61], and Mn$_{1-x}$Fe$_x$PSe$_3$ [66]. Several mechanisms [65] [61] [66] have been proposed for the suppression of $T_N$, such as the attenuation of the magnetic interaction due to disordered arrangements of the mixed metals ions with dissimilar ionic radius and outer shell electrons, and magnetic frustration among the 3d metal ions in the magnetic sites owing to competition between two different AFM structures. The strong variation of $T_N$ in heavily substituted Ni$_{1-x}$Mn$_x$PS$_3$ samples may also share similar scenarios. However, for lightly substituted samples with $x$ close to 0 or 1, $T_N$ only changes slightly as compared to the parent compounds NiPS$_3$ and MnPS$_3$. It is quite interesting to find that the light Ni-Mn substitution only weakly alters the magnetic ordering temperature but drastically suppresses the SF transitions. This suggests that the efficient suppression of the SF transition with light magnetic substitution in Ni$_{1-x}$Mn$_x$PS$_3$ can be attributed to the tuning of single ion isotropy rather than exchange interaction.

In conclusion, we have demonstrated very efficient tunning of the SF transition by light Ni-Mn substitution, which is likely attributed to single ion isotropy tuned by trigonal distortion. Such strong sensitivity suggests that magnetic substitutions can be an effective technique to control magnetism in $MPX_3$ vdWs magnetic materials, leading to a deeper understanding of low dimensional magnetism and providing insight into strategies for future magnetic device development.

**Acknowledgements**

This work is primarily suported by US Department of Energy, Office of Science, Basic Energy Sciences program under Award No. DE-SC0019467. R. B. acknowledges the support from Chancellor's Innovation and Collaboration Fund at the University of Arkansas.

**References**

[1] B. Huang, G. Clark, E. Navarro-Moratalla, D. R. Klein, R. Cheng, K. L. Seyler, D. Zhong, E. Schmidgall, M. A. McGuire, D. H. Cobden, W. Yao, D. Xiao, P. Jarillo-Herrero, and X. Xu, *Layer-Dependent Ferromagnetism in a van Der Waals Crystal down to the Monolayer Limit*, Nature **546**, 7657 (2017).

[2] C. Gong, L. Li, Z. Li, H. Ji, A. Stern, Y. Xia, T. Cao, W. Bao, C. Wang, Y. Wang, Z. Q. Qiu, R. J. Cava, S. G. Louie, J. Xia, and X. Zhang, *Discovery of Intrinsic Ferromagnetism in Two-Dimensional van Der Waals Crystals*, Nature **546**, 265 (2017).

[3] Z. Fei, B. Huang, P. Malinowski, W. Wang, T. Song, J. Sanchez, W. Yao, D. Xiao, X. Zhu, A. F. May, W. Wu, D. H. Cobden, J.-H. Chu, and X. Xu, *Two-Dimensional Itinerant Ferromagnetism in Atomically Thin $Fe_3GeTe_2$*, Nat. Mater. **17**, 9 (2018).

[4] X. Cai, T. Song, N. P. Wilson, G. Clark, M. He, X. Zhang, T. Taniguchi, K. Watanabe, W. Yao, D. Xiao, M. A. McGuire, D. H. Cobden, and X. Xu, *Atomically Thin $CrCl_3$: An In-Plane Layered Antiferromagnetic Insulator*, Nano Lett. **19**, 3993 (2019).

[5] G. Long, H. Henck, M. Gibertini, D. Dumcenco, Z. Wang, T. Taniguchi, K. Watanabe, E. Giannini, and A. F. Morpurgo, *Persistence of Magnetism in Atomically Thin $MnPS_3$ Crystals*, Nano Lett. **20**, 2452 (2020).

[6] L. Kang, C. Ye, X. Zhao, X. Zhou, J. Hu, Q. Li, D. Liu, C. M. Das, J. Yang, D. Hu, J. Chen, X. Cao, Y. Zhang, M. Xu, J. Di, D. Tian, P. Song, G. Kutty, Q. Zeng, Q. Fu, Y. Deng, J. Zhou, A. Ariando, F. Miao, G. Hong, Y. Huang, S. J. Pennycook, K.-T. Yong, W. Ji, X. Renshaw Wang, and Z. Liu, *Phase-Controllable Growth of Ultrathin 2D Magnetic FeTe Crystals*, Nat. Commun. **11**, 1 (2020).

[7] S. Albarakati, C. Tan, Z.-J. Chen, J. G. Partridge, G. Zheng, L. Farrar, E. L. H. Mayes, M. R. Field, C. Lee, Y. Wang, Y. Xiong, M. Tian, F. Xiang, A. R. Hamilton, O. A. Tretiakov, D. Culcer, Y.-J. Zhao, and L. Wang, *Antisymmetric Magnetoresistance in van Der Waals $Fe_3GeTe_2$/Graphite/$Fe_3GeTe_2$ Trilayer Heterostructures*, Sci. Adv. **5**, eaaw0409 (2019).

[8] S. Jiang, L. Li, Z. Wang, K. F. Mak, and J. Shan, *Controlling Magnetism in 2D $CrI_3$ by Electrostatic Doping*, Nat. Nanotechnol. **13**, 7 (2018).

[9] Z. Wang, T. Zhang, M. Ding, B. Dong, Y. Li, M. Chen, X. Li, J. Huang, H. Wang, X. Zhao, Y. Li, D. Li, C. Jia, L. Sun, H. Guo, Y. Ye, D. Sun, Y. Chen, T. Yang, J. Zhang, S. Ono, Z. Han, and Z. Zhang, *Electric-Field Control of Magnetism in a Few-Layered van Der Waals Ferromagnetic Semiconductor*, Nat. Nanotechnol. **13**, 7 (2018).

[10] B. Huang, G. Clark, D. R. Klein, D. MacNeill, E. Navarro-Moratalla, K. L. Seyler, N. Wilson, M. A. McGuire, D. H. Cobden, D. Xiao, W. Yao, P. Jarillo-Herrero, and X. Xu, *Electrical Control of 2D Magnetism in Bilayer $CrI_3$*, Nat. Nanotechnol. **13**, 7 (2018).

[11] Y. Deng, Y. Yu, Y. Song, J. Zhang, N. Z. Wang, Z. Sun, Y. Yi, Y. Z. Wu, S. Wu, J. Zhu, J. Wang, X. H. Chen, and Y. Zhang, *Gate-Tunable Room-Temperature Ferromagnetism in Two-Dimensional $Fe_3GeTe_2$*, Nature **563**, 7729 (2018).

[12] T. Song, X. Cai, M. W.-Y. Tu, X. Zhang, B. Huang, N. P. Wilson, K. L. Seyler, L. Zhu, T. Taniguchi, K. Watanabe, M. A. McGuire, D. H. Cobden, D. Xiao, W. Yao, and X. Xu, *Giant Tunneling Magnetoresistance in Spin-Filter van Der Waals Heterostructures*, Science **360**, 1214 (2018).

[13] Y. Wu, S. Zhang, J. Zhang, W. Wang, Y. L. Zhu, J. Hu, G. Yin, K. Wong, C. Fang, C. Wan, X. Han, Q. Shao, T. Taniguchi, K. Watanabe, J. Zang, Z. Mao, X. Zhang, and K. L. Wang,

*Néel-Type Skyrmion in WTe₂/Fe₃GeTe₂ van Der Waals Heterostructure*, Nat. Commun. **11**, 1 (2020).

[14] D. R. Klein, D. MacNeill, J. L. Lado, D. Soriano, E. Navarro-Moratalla, K. Watanabe, T. Taniguchi, S. Manni, P. Canfield, J. Fernández-Rossier, and P. Jarillo-Herrero, *Probing Magnetism in 2D van Der Waals Crystalline Insulators via Electron Tunneling*, Science **360**, 1218 (2018).

[15] J. Shang, X. Tang, X. Tan, A. Du, T. Liao, S. C. Smith, Y. Gu, C. Li, and L. Kou, *Stacking-Dependent Interlayer Magnetic Coupling in 2D CrI₃/CrGeTe₃ Nanostructures for Spintronics*, ACS Appl. Nano Mater. **3**, 1282 (2020).

[16] D. Zhong, K. L. Seyler, X. Linpeng, R. Cheng, N. Sivadas, B. Huang, E. Schmidgall, T. Taniguchi, K. Watanabe, M. A. McGuire, W. Yao, D. Xiao, K.-M. C. Fu, and X. Xu, *Van Der Waals Engineering of Ferromagnetic Semiconductor Heterostructures for Spin and Valleytronics*, Sci. Adv. **3**, e1603113 (2017).

[17] Z. Wang, I. Gutiérrez-Lezama, N. Ubrig, M. Kroner, M. Gibertini, T. Taniguchi, K. Watanabe, A. Imamoğlu, E. Giannini, and A. F. Morpurgo, *Very Large Tunneling Magnetoresistance in Layered Magnetic Semiconductor CrI₃*, Nat. Commun. **9**, 1 (2018).

[18] T. Song, M. W.-Y. Tu, C. Carnahan, X. Cai, T. Taniguchi, K. Watanabe, M. A. McGuire, D. H. Cobden, D. Xiao, W. Yao, and X. Xu, *Voltage Control of a van Der Waals Spin-Filter Magnetic Tunnel Junction*, Nano Lett. **19**, 915 (2019).

[19] G. Ouvrard, R. Brec, and J. Rouxel, *Structural Determination of Some MPS₃ Layered Phases (M = Mn, Fe, Co, Ni and Cd)*, Mater. Res. Bull. **20**, 1181 (1985).

[20] W. Klingen, G. Eulenberger, and H. Hahn, *About the crystal structures of Fe₂P₂Se₆ and Fe₂P₂S₆*, Z. Für Anorg. Allg. Chem. **401**, 97 (1973).

[21] K. Du, X. Wang, Y. Liu, P. Hu, M. I. B. Utama, C. K. Gan, Q. Xiong, and C. Kloc, *Weak Van Der Waals Stacking, Wide-Range Band Gap, and Raman Study on Ultrathin Layers of Metal Phosphorus Trichalcogenides*, ACS Nano **10**, 1738 (2016).

[22] F. Wang, T. A. Shifa, P. Yu, P. He, Y. Liu, F. Wang, Z. Wang, X. Zhan, X. Lou, F. Xia, and J. He, *New Frontiers on van Der Waals Layered Metal Phosphorous Trichalcogenides*, Adv. Funct. Mater. **28**, 1802151 (2018).

[23] R. Brec, D. M. Schleich, G. Ouvrard, A. Louisy, and J. Rouxel, *Physical Properties of Lithium Intercalation Compounds of the Layered Transition-Metal Chalcogenophosphites*, Inorg. Chem. **18**, 1814 (1979).

[24] P. J. S. Foot, T. Katz, S. N. Patel, B. A. Nevett, A. R. Pieecy, and A. A. Balchin, *The Structures and Conduction Mechanisms of Lithium-Intercalated and Lithium-Substituted Nickel Phosphorus Trisulphide ($NiPS_3$), and the Use of the Material as a Secondary Battery Electrode*, Phys. Status Solidi A **100**, 11 (1987).

[25] L. Silipigni, L. Schirò, T. Quattrone, V. Grasso, G. Salvato, L. Monsù Scolaro, and G. De Luca, *Dielectric Spectra of Manganese Thiophosphate Intercalated with Sodium Ions*, J. Appl. Phys. **105**, 123703 (2009).

[26] G. Giunta, V. Grasso, F. Neri, and L. Silipigni, *Electrical Conductivity of Lithium-Intercalated Thiophosphate $NiPS_3$ Single Crystals*, Phys. Rev. B **50**, 8189 (1994).

[27] L. Silipigni, C. Calareso, G. M. Curró, F. Neri, V. Grasso, H. Berger, G. Margaritondo, and R. Ponterio, *Effects of Lithium Intercalation on the Electronic Properties of $FePS_3$ Single Crystals*, Phys. Rev. B **53**, 13928 (1996).

[28] C.-T. Kuo, M. Neumann, K. Balamurugan, H. J. Park, S. Kang, H. W. Shiu, J. H. Kang, B. H. Hong, M. Han, T. W. Noh, and J.-G. Park, *Exfoliation and Raman Spectroscopic Fingerprint of Few-Layer NiPS$_3$ Van Der Waals Crystals*, Sci. Rep. **6**, 1 (2016).

[29] D. H. Luong, T. L. Phan, G. Ghimire, D. L. Duong, and Y. H. Lee, *Revealing Antiferromagnetic Transition of van Der Waals MnPS$_3$ via Vertical Tunneling Electrical Resistance Measurement*, APL Mater. **7**, 081102 (2019).

[30] S. Lee, K.-Y. Choi, S. Lee, B. H. Park, and J.-G. Park, *Tunneling Transport of Mono- and Few-Layers Magnetic van Der Waals MnPS$_3$*, APL Mater. **4**, 086108 (2016).

[31] H. Chu, C. J. Roh, J. O. Island, C. Li, S. Lee, J. Chen, J.-G. Park, A. F. Young, J. S. Lee, and D. Hsieh, *Linear Magnetoelectric Phase in Ultrathin MnPS$_3$ Probed by Optical Second Harmonic Generation*, Phys. Rev. Lett. **124**, 027601 (2020).

[32] X. Wang, J. Cao, Z. Lu, A. Cohen, H. Kitadai, T. Li, M. Wilson, C. H. Lui, D. Smirnov, S. Sharifzadeh, and X. Ling, *Spin-Induced Linear Polarization of Excitonic Emission in Antiferromagnetic van Der Waals Crystals*, ArXiv200607952 Cond-Mat (2020).

[33] S. N. Neal, H.-S. Kim, K. A. Smith, A. V. Haglund, D. G. Mandrus, H. A. Bechtel, G. L. Carr, K. Haule, D. Vanderbilt, and J. L. Musfeldt, *Near-Field Infrared Spectroscopy of Monolayer MnPS$_3$*, Phys. Rev. B **100**, 075428 (2019).

[34] G. Long, T. Zhang, X. Cai, J. Hu, C. Cho, S. Xu, J. Shen, Z. Wu, T. Han, J. Lin, J. Wang, Y. Cai, R. Lortz, Z. Mao, and N. Wang, *Isolation and Characterization of Few-Layer Manganese Thiophosphite*, ACS Nano **11**, 11330 (2017).

[35] J.-U. Lee, S. Lee, J. H. Ryoo, S. Kang, T. Y. Kim, P. Kim, C.-H. Park, J.-G. Park, and H. Cheong, *Ising-Type Magnetic Ordering in Atomically Thin FePS$_3$*, Nano Lett. **16**, 7433 (2016).

[36] K. Kim, S. Y. Lim, J.-U. Lee, S. Lee, T. Y. Kim, K. Park, G. S. Jeon, C.-H. Park, J.-G. Park, and H. Cheong, *Suppression of Magnetic Ordering in XXZ-Type Antiferromagnetic Monolayer NiPS$_3$*, Nat. Commun. **10**, 1 (2019).

[37] P. A. Joy and S. Vasudevan, *Magnetism in the Layered Transition-Metal Thiophosphates MPS$_3$ (M=Mn, Fe, and Ni)*, Phys. Rev. B **46**, 5425 (1992).

[38] A. R. Wildes, V. Simonet, E. Ressouche, G. J. Mcintyre, M. Avdeev, E. Suard, S. A. Kimber, D. Lançon, G. Pepe, and B. Moubaraki, *Magnetic Structure of the Quasi-Two-Dimensional Antiferromagnet NiPS$_3$*, Phys. Rev. B **92**, 224408 (2015).

[39] K. C. Rule, G. J. McIntyre, S. J. Kennedy, and T. J. Hicks, *Single-Crystal and Powder Neutron Diffraction Experiments on FePS$_3$: Search for the Magnetic Structure*, Phys. Rev. B **76**, 134402 (2007).

[40] A. R. Wildes, H. M. Rønnow, B. Roessli, M. J. Harris, and K. W. Godfrey, *Static and Dynamic Critical Properties of the Quasi-Two-Dimensional Antiferromagnet MnPS$_3$*, Phys. Rev. B **74**, 094422 (2006).

[41] D. Lançon, H. C. Walker, E. Ressouche, B. Ouladdiaf, K. C. Rule, G. J. McIntyre, T. J. Hicks, H. M. Rønnow, and A. R. Wildes, *Magnetic Structure and Magnon Dynamics of the Quasi-Two-Dimensional Antiferromagnet FePS$_3$*, Phys. Rev. B **94**, 214407 (2016).

[42] A. R. Wildes, B. Roessli, B. Lebech, and K. W. Godfrey, *Spin Waves and the Critical Behaviour of the Magnetization in MnPS3*, J. Phys. Condens. Matter **10**, 6417 (1998).

[43] A. R. Wildes, V. Simonet, E. Ressouche, R. Ballou, and G. J. McIntyre, *The Magnetic Properties and Structure of the Quasi-Two-Dimensional Antiferromagnet CoPS$_3$*, J. Phys. Condens. Matter **29**, 455801 (2017).

[44] T. Sekine, M. Jouanne, C. Julien, and M. Balkanski, *Light-Scattering Study of Dynamical Behavior of Antiferromagnetic Spins in the Layered Magnetic Semiconductor FePS$_3$*, Phys. Rev. B **42**, 8382 (1990).

[45] Y. Takano, N. Arai, A. Arai, Y. Takahashi, K. Takase, and K. Sekizawa, *Magnetic Properties and Specific Heat of MPS$_3$ (M=Mn, Fe, Zn)*, J. Magn. Magn. Mater. **272–276**, E593 (2004).

[46] S. Y. Kim, T. Y. Kim, L. J. Sandilands, S. Sinn, M.-C. Lee, J. Son, S. Lee, K.-Y. Choi, W. Kim, B.-G. Park, C. Jeon, H.-D. Kim, C.-H. Park, J.-G. Park, S. J. Moon, and T. W. Noh, *Charge-Spin Correlation in van Der Waals Antiferromagnet NiPS$_3$*, Phys. Rev. Lett. **120**, 136402 (2018).

[47] H.-S. Kim, K. Haule, and D. Vanderbilt, *Mott Metal-Insulator Transitions in Pressurized Layered Trichalcogenides*, Phys. Rev. Lett. **123**, 236401 (2019).

[48] Y. Wang, Z. Zhou, T. Wen, Y. Zhou, N. Li, F. Han, Y. Xiao, P. Chow, J. Sun, M. Pravica, A. L. Cornelius, W. Yang, and Y. Zhao, *Pressure-Driven Cooperative Spin-Crossover, Large-Volume Collapse, and Semiconductor-to-Metal Transition in Manganese (II) Honeycomb Lattices*, J. Am. Chem. Soc. **138**, 15751 (2016).

[49] C. R. S. Haines, M. J. Coak, A. R. Wildes, G. I. Lampronti, C. Liu, P. Nahai-Williamson, H. Hamidov, D. Daisenberger, and S. S. Saxena, *Pressure-Induced Electronic and Structural Phase Evolution in the van Der Waals Compound FePS$_3$*, Phys. Rev. Lett. **121**, 266801 (2018).

[50] M. Tsurubayashi, K. Kodama, M. Kano, K. Ishigaki, Y. Uwatoko, T. Watanabe, K. Takase, and Y. Takano, *Metal-Insulator Transition in Mott-Insulator FePS$_3$*, AIP Adv. **8**, 101307 (2018).

[51] M. J. Coak, D. M. Jarvis, H. Hamidov, C. R. S. Haines, P. L. Alireza, C. Liu, S. Son, I. Hwang, G. I. Lampronti, D. Daisenberger, P. Nahai-Williamson, A. R. Wildes, S. S. Saxena, and J.-G. Park, *Tuning Dimensionality in Van-Der-Waals Antiferromagnetic Mott Insulators TMPS$_3$*, J. Phys. Condens. Matter **32**, 124003 (2019).

[52] R. A. Evarestov and A. Kuzmin, *Origin of Pressure-Induced Insulator-to-Metal Transition in the van Der Waals Compound FePS$_3$ from First-Principles Calculations*, J. Comput. Chem. **41**, 1337 (2020).

[53] X. Ma, Y. Wang, Y. Yin, B. Yue, J. Dai, J. Ji, F. Jin, F. Hong, J.-T. Wang, Q. Zhang, and X. Yu, *Dimensional Crossover Tuned by Pressure in Layered Magnetic NiPS$_3$*, ArXiv200914051 Cond-Mat (2020).

[54] M. J. Coak, S. Son, D. Daisenberger, H. Hamidov, C. R. S. Haines, P. L. Alireza, A. R. Wildes, C. Liu, S. S. Saxena, and J.-G. Park, *Isostructural Mott Transition in 2D Honeycomb Antiferromagnet V$_{0.9}$PS$_3$*, Npj Quantum Mater. **4**, 1 (2019).

[55] Y. Wang, J. Ying, Z. Zhou, J. Sun, T. Wen, Y. Zhou, N. Li, Q. Zhang, F. Han, Y. Xiao, P. Chow, W. Yang, V. V. Struzhkin, Y. Zhao, and H. Mao, *Emergent Superconductivity in an Iron-Based Honeycomb Lattice Initiated by Pressure-Driven Spin-Crossover*, Nat. Commun. **9**, 1 (2018).

[56] N. Chandrasekharan and S. Vasudevan, *Dilution of a Layered Antiferromagnet: Magnetism in Mn$_x$Zn$_{1-x}$PS$_3$*, Phys. Rev. B **54**, 14903 (1996).

[57] D. J. Goossens, A. J. Studer, S. J. Kennedy, and T. J. Hicks, *The Impact of Magnetic Dilution on Magnetic Order in MnPS3*, J. Phys. Condens. Matter **12**, 4233 (2000).

[58] A. M. Mulders, J. C. P. Klaasse, D. J. Goossens, J. Chadwick, and T. J. Hicks, *High-Field Magnetization in the Diluted Quasi-Two-Dimensional Heisenberg Antiferromagnet $Mn_{1-x}Zn_xPS_3$*, J. Phys. Condens. Matter **14**, 8697 (2002).

[59] Y. Takano, A. Arai, Y. Takahashi, K. Takase, and K. Sekizawa, *Magnetic Properties and Specific Heat of New Spin Glass $Mn_{0.5}Fe0.5PS_3$*, J. Appl. Phys. **93**, 8197 (2003).

[60] J. N. Graham, M. J. Coak, S. Son, E. Suard, J.-G. Park, L. Clark, and A. R. Wildes, *Local Nuclear and Magnetic Order in the Two-Dimensional Spin Glass $Mn_{0.5}Fe_{0.5}PS_3$*, Phys. Rev. Mater. **4**, 084401 (2020).

[61] T. Masubuchi, H. Hoya, T. Watanabe, Y. Takahashi, S. Ban, N. Ohkubo, K. Takase, and Y. Takano, *Phase Diagram, Magnetic Properties and Specific Heat of $Mn_{1-x}Fe_xPS_3$*, J. Alloys Compd. **460**, 668 (2008).

[62] V. Manríquez, P. Barahona, and O. Peña, *Physical Properties of the Cation-Mixed M′MPS$_3$ Phases*, Mater. Res. Bull. **35**, 1889 (2000).

[63] D. J. Goossens, S. Brazier-Hollins, D. R. James, W. D. Hutchison, and J. R. Hester, *Magnetic Structure and Glassiness in $Fe_{0.5}Ni_{0.5}PS_3$*, J. Magn. Magn. Mater. **334**, 82 (2013).

[64] Y. He, Y.-D. Dai, H. Huang, J. Lin, and Y. Hsia, *The Ordering Distribution of the Metal Ions in the Layered Cation-Mixed Phosphorus Trisulfides $Mn_xFe_{1-x}PS_3$*, J. Alloys Compd. **359**, 41 (2003).

[65] D. J. Goossens and T. J. Hicks, *The Magnetic Phase Diagram of $Mn_xZn_{1-x}PS_3$*, J. Phys. Condens. Matter **10**, 7643 (1998).

[66] A. Bhutani, J. L. Zuo, R. D. McAuliffe, C. R. dela Cruz, and D. P. Shoemaker, *Strong Anisotropy in the Mixed Antiferromagnetic System $Mn_{1-x}Fe_xPSe_3$*, Phys. Rev. Mater. **4**, 034411 (2020).

[67] B. L. Chittari, Y. Park, D. Lee, M. Han, A. H. MacDonald, E. Hwang, and J. Jung, *Electronic and Magnetic Properties of Single-Layer MPX$_3$ Metal Phosphorous Trichalcogenides*, Phys. Rev. B **94**, 184428 (2016).

[68] K. Okuda, K. Kurosawa, S. Saito, M. Honda, Z. Yu, and M. Date, *Magnetic Properties of Layered Compound MnPS$_3$*, J. Phys. Soc. Jpn. **55**, 4456 (1986).

[69] D. J. Goossens, A. R. Wildes, C. Ritter, and T. J. Hicks, *Ordering and the Nature of the Spin Flop Phase Transition in MnPS$_3$*, J. Phys. Condens. Matter **12**, 1845 (2000).

[70] A. R. Wildes, D. Lançon, M. K. Chan, F. Weickert, N. Harrison, V. Simonet, M. E. Zhitomirsky, M. V. Gvozdikova, T. Ziman, and H. M. Rønnow, *High Field Magnetization of FePS$_3$*, Phys. Rev. B **101**, 024415 (2020).

[71] A. P. Dioguardi, S. Selter, U. Peeck, S. Aswartham, M.-I. Sturza, R. Murugesan, M. S. Eldeeb, L. Hozoi, B. Büchner, and H.-J. Grafe, *Quasi-Two-Dimensional Magnetic Correlations in Ni$_2$P$_2$S$_6$ Probed by $^{31}$P NMR*, Phys. Rev. B **102**, 064429 (2020).

[72] F. B. Anderson and H. B. Callen, *Statistical Mechanics and Field-Induced Phase Transitions of the Heisenberg Antiferromagnet*, Phys. Rev. **136**, A1068 (1964).

[73] L. J. D. Jongh and A. R. Miedema, *Experiments on Simple Magnetic Model Systems*, Adv. Phys. **50**, 947 (2001).

[74] E. Emmanouilidou, J. Liu, D. Graf, H. Cao, and N. Ni, *Spin-Flop Phase Transition in the Orthorhombic Antiferromagnetic Topological Semimetal Cu$_{0.95}$MnAs*, J. Magn. Magn. Mater. **469**, 570 (2019).

[75] G. Gitgeatpong, M. Suewattana, S. Zhang, A. Miyake, M. Tokunaga, P. Chanlert, N. Kurita, H. Tanaka, T. J. Sato, Y. Zhao, and K. Matan, *High-Field Magnetization and Magnetic Phase Diagram of α−Cu$_2$V$_2$O$_7$*, Phys. Rev. B **95**, 245119 (2017).

[76] S. M. Wu, W. Zhang, A. KC, P. Borisov, J. E. Pearson, J. S. Jiang, D. Lederman, A. Hoffmann, and A. Bhattacharya, *Antiferromagnetic Spin Seebeck Effect*, Phys. Rev. Lett. **116**, 097204 (2016).

[77] D. Li, Z. Han, J. G. Zheng, X. L. Wang, D. Y. Geng, J. Li, and Z. D. Zhang, *Spin Canting and Spin-Flop Transition in Antiferromagnetic $Cr_2O_3$ Nanocrystals*, J. Appl. Phys. **106**, 053913 (2009).

[78] D. Tobia, E. Winkler, R. D. Zysler, M. Granada, and H. E. Troiani, *Size Dependence of the Magnetic Properties of Antiferromagnetic $Cr_2O_3$ Nanoparticles*, Phys. Rev. B **78**, 104412 (2008).

[79] W. Zhang, K. Nadeem, H. Xiao, R. Yang, B. Xu, H. Yang, and X. G. Qiu, *Spin-Flop Transition and Magnetic Phase Diagram in $CaCo_2As_2$ Revealed by Torque Measurements*, Phys. Rev. B **92**, 144416 (2015).

[80] C. Kim, J. Jeong, P. Park, T. Masuda, S. Asai, S. Itoh, H.-S. Kim, A. Wildes, and J.-G. Park, *Spin Waves in the Two-Dimensional Honeycomb Lattice XXZ-Type van Der Waals Antiferromagnet $CoPS_3$*, Phys. Rev. B **102**, 184429 (2020).

[81] D. Lançon, R. A. Ewings, T. Guidi, F. Formisano, and A. R. Wildes, *Magnetic Exchange Parameters and Anisotropy of the Quasi-Two-Dimensional Antiferromagnet $NiPS_3$*, Phys. Rev. B **98**, 134414 (2018).

[82] D. J. Goossens, *Dipolar Anisotropy in Quasi-2D Honeycomb Antiferromagnet $MnPS_3$*, Eur. Phys. J. B **78**, 305 (2010).

[83] Y. Gu, Q. Zhang, C. Le, Y. Li, T. Xiang, and J. Hu, *Ni-Based Transition Metal Trichalcogenide Monolayer: A Strongly Correlated Quadruple-Layer Graphene*, Phys. Rev. B **100**, 165405 (2019).

**Figures**

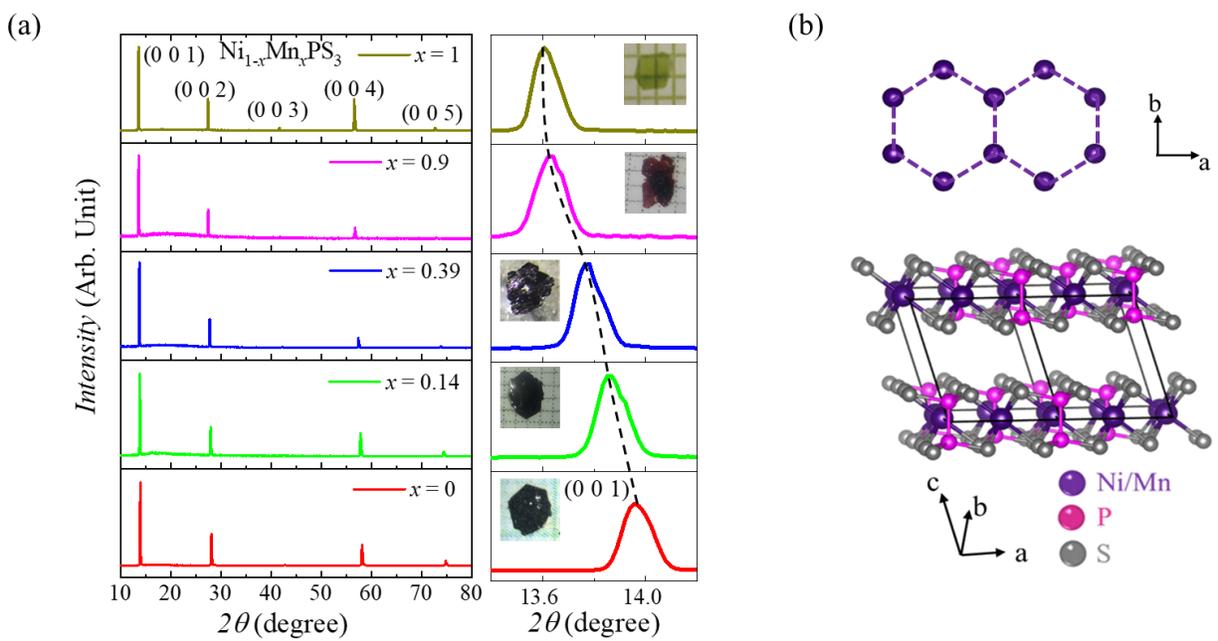

**FIG. 1.** (a) Single-crystal x-ray-diffraction pattern for $Ni_{1-x}Mn_xPS_3$ showing the (00$L$) reflections. Right panels show (001) diffraction peak. Inset: Optical microscope images of the single crystals. (b) Crystal structure of $MPS_3$ ($M$ = Ni/Mn).

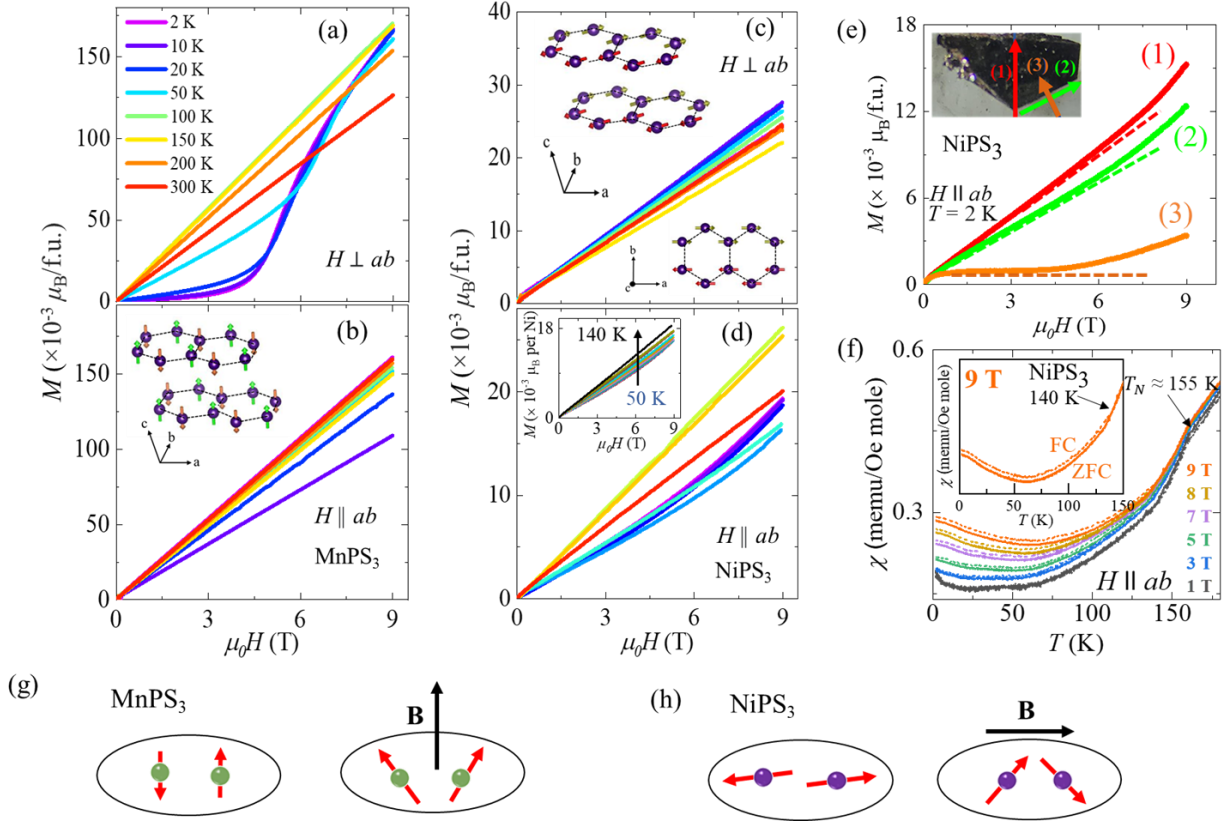

**FIG. 2.** Field induced spin flop transitions in pristine MnPS$_3$ and NiPS$_3$. (a-b) Isothermal magnetization of MnPS$_3$ at different temperatures under (a) $H \perp ab$ and (b) $H \| ab$. Inset in (b): magnetic structure of MnPS$_3$. (c-d) Isothermal magnetization of NiPS$_3$ at different temperatures under (a) $H \perp ab$ and (d) $H \| ab$. Upper inset in (c): 3D view of magnetic structure of NiPS$_3$. Lower inset in (c): Top view of magnetic structure of NiPS$_3$. Inset in (d): Isothermal magnetization at 50 – 140 K measured with $H \| ab$. The same color code is used indicate temperatures for (a-d). (e) Isothermal magnetization at 2 K under $H \| ab$ with different in-plane field orientations. Inset: Optical microscope image of NiPS$_3$ single crystal with arrows pointing the applied field direction. (f) Temperature dependence of susceptibility of NiPS$_3$ measure with 1, 3, 5, 7, 8, and 9 T fields, measured with $H \| ab$. Inset: Temperature dependence of susceptibility at 9 T under $H \| ab$. The solid and doted lines represent zero-field cooled (ZFC) and field-cooled (FC) data, respectively.

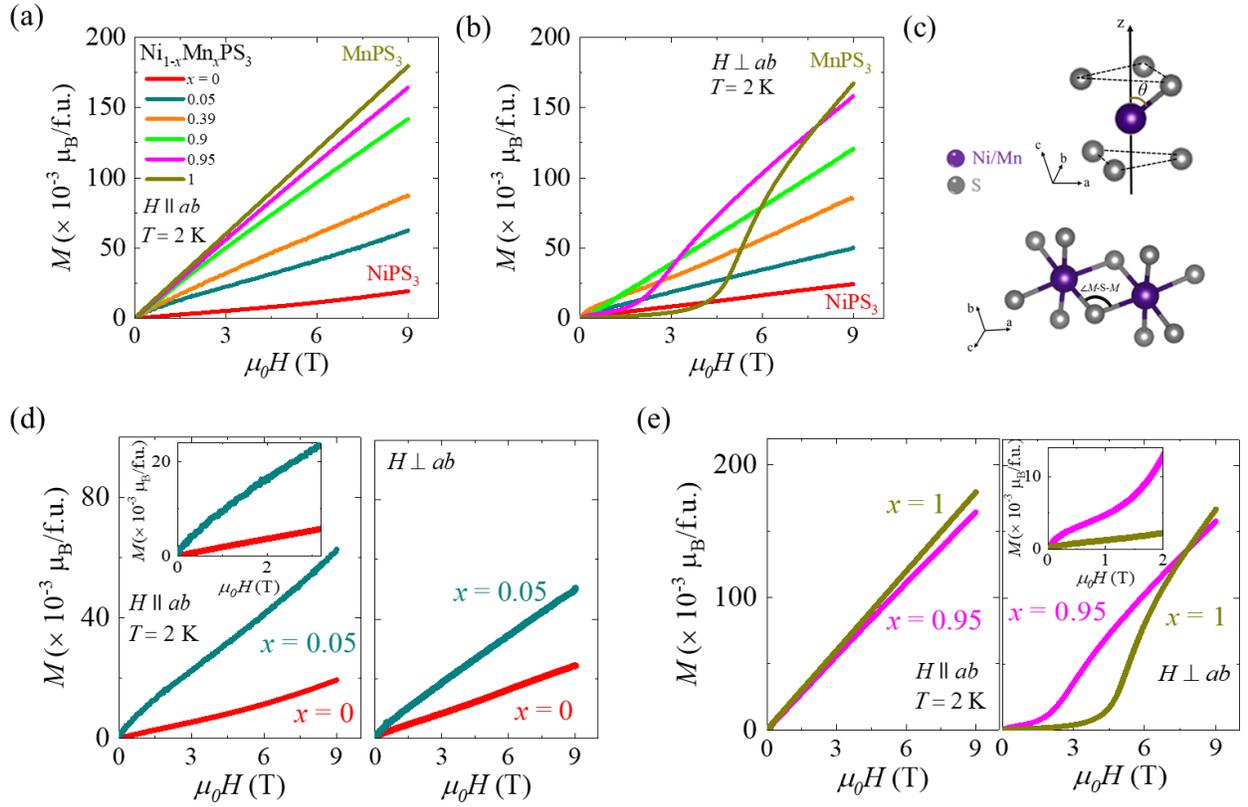

**FIG. 3.** (a-b) Isothermal magnetization of $Ni_{1-x}Mn_xPS_3$ samples ($0 \leq x \leq 1$) at $T = 2$ K under (a) $H \| ab$ and (b) $H \perp ab$. The same color code is used to indicate temperatures for (a) and (b). (c) Trigonal distortion in $M$PS$_3$ compounds. $z$-axis is the trigonal axis. (d) Isothermal magnetization of $Ni_{1-x}Mn_xPS_3$ ($x = 0$ and 0.05) at $T = 2$ K under $H \| ab$ (Left panel) and $H \perp ab$ (Right panel). Inset: zoom in of the low-field magnetization. (e) Isothermal magnetization of $Ni_{1-x}Mn_xPS_3$ ($x = 0.95$ and 1) at $T = 2$ K under $H \| ab$ (Left panel) and $H \perp ab$ (Right panel). Inset: zoom in of the low-field magnetization.

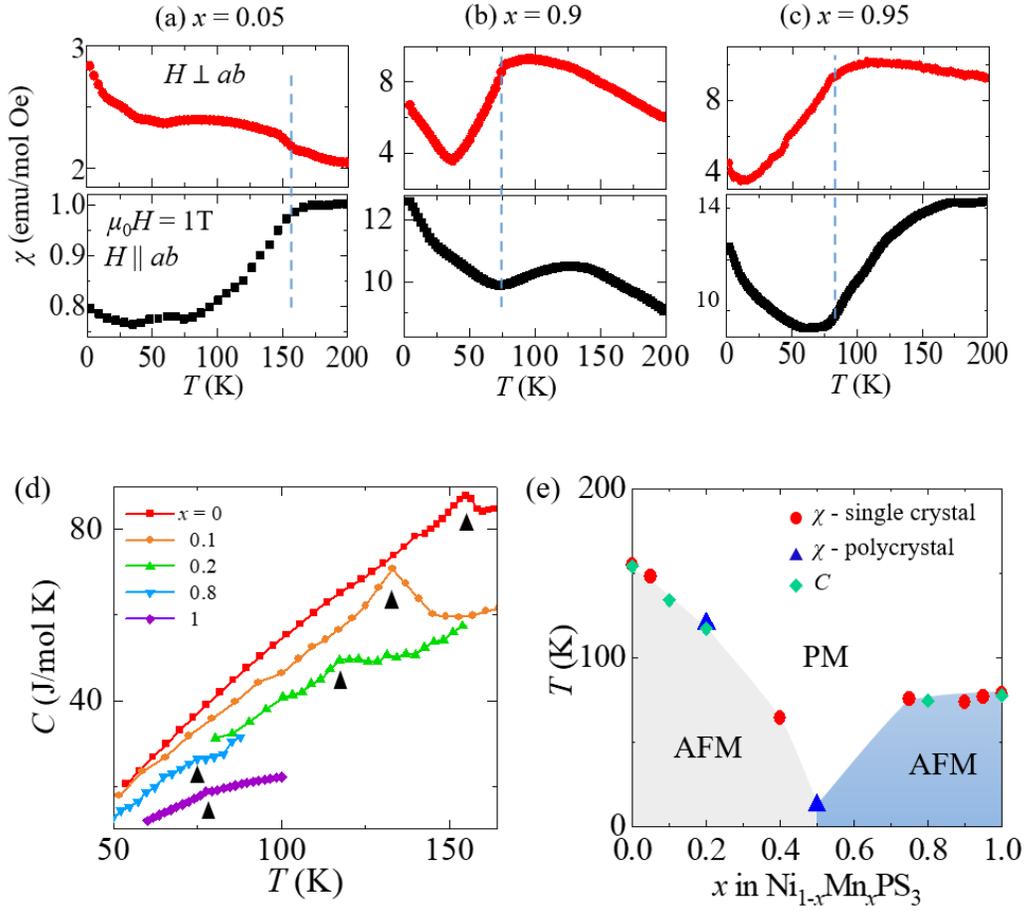

**FIG. 4.** (a-c) Temperature dependence of out-of-plane ($H\perp ab$, red, upper panels) and in-plane ($H\|ab$, black, lower panels) susceptibilities of (a) $Ni_{0.95}Mn_{0.05}PS_3$, (b) $Ni_{0.1}Mn_{0.9}PS_3$, and (c) $Ni_{0.05}Mn_{0.95}PS_3$ samples. The dashed lines denote $T_N$. (d) Temperature dependence of heat capacity of $Ni_{1-x}Mn_xPS_3$ samples with $x = 0, 0.1, 0.2, 0.8$ and 1. The black triangles denote $T_N$. Data are shifted for clarity. (e) Magnetic phase diagram of $Ni_{1-x}Mn_xPS_3$ ($0 \leq x \leq 1$). The transition temperatures are determined by susceptibility measurements on single- ($\chi$-single crystal) and poly-crystals ($\chi$-polycrystal), and heat capacity ($C$).